\documentclass[conference]{IEEEtran}
\IEEEoverridecommandlockouts

\usepackage{cite}
\usepackage{amsmath,amssymb,amsfonts}
\usepackage{algorithmic}
\usepackage{graphicx}
\usepackage{textcomp}
\usepackage{xcolor}
\def\BibTeX{{\rm B\kern-.05em{\sc i\kern-.025em b}\kern-.08em
    T\kern-.1667em\lower.7ex\hbox{E}\kern-.125emX}}
\begin{document}

\title{Verification and Validation of Autonomous Systems\\
}

\author{\IEEEauthorblockN{1\textsuperscript{st} Sneha Sudhir Shetiya}
\IEEEauthorblockA{\textit{IEEE Senior Member.)} \\
\textit{Michigan, USA} \\
sneha.shetiya@ieee.org}
\and
\IEEEauthorblockN{2\textsuperscript{nd} Vikas Vyas}
\IEEEauthorblockA{\textit{IEEE Senior Member} \\
\textit{California, USA}\\
vikas.vyas@mercedes-benz.com}
\and
\IEEEauthorblockN{3\textsuperscript{th} Shreyas Renukuntla}
\IEEEauthorblockA{\textit{PhD Student} \\
\textit{Oakland University}\\
Michigan, USA \\
shreyas.renukuntla@gmail.com}
}

\maketitle

\begin{abstract}
This paper describes how to proficiently prevent software defects in autonomous vehicles, discover and correct defects if they are encountered, and create a higher level of assurance in the software product development phase. It also describes how to ensure high assurance on software reliability.
\end{abstract}

\begin{IEEEkeywords}
Software Development V-cycle, Hardware in Loop(HIL), Software in Loop(SIL), Vehicle in Loop(VIL), Final Product Acceptance Test(FPAT)
\end{IEEEkeywords}

\section{Introduction}
In today’s world software is pervasive. Software helps make life convenient. It controls our home appliances, automobiles, phones, and entertainment. It increases our productivity at work, speeds up our communication, and improves our medical care.  It affects nearly every aspect of modern life. Software is the backbone of modern civilization, driving innovation and progress across industries. Whatever the future holds, software will remain a crucial part of it. Software is getting more complex due to many reasons, such as an increase in software application diversity, an increase in software platform types, and an increase reliability on “third-party” software. As a consequence, it is crucial to produce reliable software. Software failure often may mean that some entertainment application is not entertaining as intended, or it could result in a life-or-death situation in a hospital or a mass transit system.

\begin{figure}[htbp]
\centerline{\includegraphics{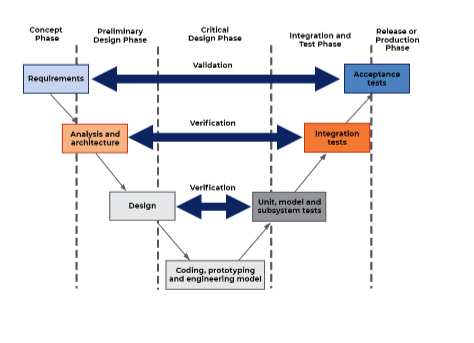}}
\caption{V-cycle Software model}
\label{fig1}
\end{figure}

\subsection{Phases of Software Product Development}
Generally, the V-model is considered for software development and testing methodology. The V-model is considered a variant of the traditional Waterfall approach. Besides the standard Waterfall methodology’s life cycle phases, which overflow one after the other in a linear development process. The V-model is categorized by a testing phase running in parallel to each development phase.
The advance of the Waterfall methodology to integrate more distinct testing phases is described by the V-shaped infographic \ref{fig1} which explains the advanced approach. The V-model represents the extended ‘Verification and Validation model’.
The book Developing and Managing Embedded Systems and Products – Methods, Techniques, Tools, Processes, and Teamwork by Kim R. Fowler and Craig L.\cite{b1} Silver defines the V-model’s verification and validation components as follows:“Verification is an objective set of tests to confirm that the product meets the metrics of the requirements, while validation seeks to demonstrate that the product meets the original intent.”

\subsection{Steps of the V-cycle}
\subsubsection{Requirements}
Gathering the stakeholder requirements is essential for any feature implementation in a software module.These requirements can be external to the Organization driven from the Original Equipment Manufacturer(OEM) or internal from the product team. These requirements are essential as they form the basis for the next steps in the process.\\

\subsubsection{Analysis and Architecture}
This step involves the system and the software architecture. In other versions, it is also broken down into subsystem or component architecture. Tools like Cameo\cite{b2} and IBM Rhapsody \cite{b3} are used for modeling subsystems. Standardized tools are essential for mapping to requirements and code.\\

\subsubsection{Design}
The system and software design phase involve designing all the sub -functions to realize the architecture. The functions should be detailed to start with unit design and coding in the next step.\\

\subsubsection{Coding and prototyping}
This step is completely the expertise of software developers which involves writing code to realize the feature.\\

\subsubsection{Unit, model and subsystem tests}
As we transition to the right side of the V-cycle, bottom up testing of the implemented feature is focused upon. The current diagram \ref{fig1}displays unit, model and subsystem tests in one bug block. These can be broken down into three additional blocks. In production implementation, we have unit tests for code which are done mostly on developer side. Model based tests can involve code blocks from multiple teams to check particular functionality. Finally the subsystem test , the design for which we defined on the left side of the V-cycle.\\

\subsubsection{Integration tests}
The integration tests are done at multiple levels depending on the platform. The three major types in Automotive are Motor in Loop(MIL), Software in Loop(SIL) and Hardware in Loop(HIL). These three tests are performed using different set of tools. For MIL, there are dedicated Matlab tools like TargetLink\cite{b4}, for SIL there are tools and for HIL, its essentially hardware on bench.
Some good quality HILs existing today are Vector HILs, the ETAS HIL\cite{b5}\\

\subsubsection{Acceptance tests}
These are broken into Alpha and Beta tests. Alpha tests are for consumers internal to the Organization and Beta tests are for consumers external to the Organization.
Every software roll out has a beta test associated to it. Sometimes alpha tests are missed. But beta tests are essentially carried out to check the confidence of the general public in the functionality. One good example is Tesla's  self-driving feature\cite{b6} which has been rolled out in the United States as a Beta version and is accompanied by many updates to it as well.

\subsection{V-cycle vs the traditional waterfall model}
Traditional waterfall model did not have the scope to interact between different steps from Safety goal to Beta testing phase.
The v-cycle model has those capabilities where is interaction between every phase on the left side of V-cycle to the right. This is the major advantage why all the product companies have adopted V-cycle model for development. These connections majorly help in fixing defects identified on the right side as they are communicated to the left side upon identification. This is for quicker turnaround of fixing defects in the software development cycle. This was not possible in the waterfall model. 
The final connection between Requirements and Acceptance testing is validation. The intermediate connections at earlier phases come under the umbrella of verification.

\section{Software components of Autonomous Systems}
On a higher level, few major components that the software stack for Autonomous vehicles consist of is follows:

\subsection{Perception}
Perception as the name suggests is the objects to be perceived in the environment in the set Operational Design Domain(ODD)\cite{b7}. 
This is done through the visual sensor that are part of the sensor, stack i.e., the cameras. Based on the position in the ego vehicle , the dimensions of the camera may change. Nowadays 8 mega pixel cameras are mostly used for rear and side view perception models. The front is usually a fish eyed lens as its also used for calibration and broader capture of the scene.
\subsection{Localization and mapping}
\subsubsection{Localization} Once the objects are perceived, we need to sense and place them on the available maps. Maps are again made available through raw sensor data and collection of images across pre defined routes. Notably, sensed/perceived data during an Autonomous Vehicle(AV) trip is typically also used to update or influence the state of the mapping database.
\subsubsection{Mapping} Here maps are the inputs to assist generation of global routes and influence behavioral/local plans.
\subsection{Prediction}
This focuses on three potential actions of drivers of the AV:
\begin{itemize}
\item  Infer what other drivers want to do in the future.

\item Infer with what traits the driver’s goal will be pursued.

\item Predict the future states of each traffic participant.
\end{itemize}
Currently, the most promising prediction/behavior estimation approaches utilize data-driven learning techniques. The learning methods supporting these techniques will be further discussed in the simulation section of this article.
\subsection{Planning and Control}
Planning will produce elaborate driving trajectories by taking into account the high level instructions from the global planner, specific maneuvers requested by the behavior planner, as well as physical limitations of the vehicle.

\subsubsection{Important challenges in local planning}
\begin{itemize}
 \item Trajectory Generation:
The search space is immense. Therefore, computationally this is a costly task to find the optimal trajectories. Heuristic-based methods\cite{b8}, variational planning for instance, are very often used for efficient exploration of the solution space.
Another approach to generating diverse trajectories utilizes sampling-based and learning-based techniques.
\item Trajectory Quality Evaluation:
A suitable objective function to measure the quality of trajectory generation is quite difficult to define.
\end{itemize}
Safety, legality, comfort, efficiency, and compliance with traffic rules are commonly used evaluation criteria. The factors are often difficult to quantify and balance, such as perceived safety and comfort.
Thus, the local planner is needed in order to ensure the safe and efficient operation of autonomous vehicles, all these challenges must be addressed for their widespread deployment.
\section{Simulation}
Simulation plays a critical role in the development of Autonomous Vehicles (AVs) since it allows the development of algorithms in a controlled environment, thereby allowing training and evaluation in a realistic scenario. Most importantly, a realistic and diverse driving scenario helps bridge the problem of real-world data since some events happen less frequently.
  Some of the key benefits of using simulation include:
  \begin{itemize}
      \item Data Augmentation: Developing diverse and challenging scenarios which improves robustness.
\item Accelerated Learning: Training models faster by generating large amounts of synthetic data.
\item Model Architecture Exploration: Testing various model architectures, performing hyper-parameter tuning, to find the best configurations.
\item Safe and Controlled Testing: Simulate AV behavior in different scenarios without posing a risk of accidents in the real world.
  \end{itemize}
Leverage simulation to optimize developers to improve safety, reliability, and performance of self-driving vehicles.

\section{Calibration testing and challenges}
One important aspect often ignored in Autonomous vehicles software development is to check the accuracy of calibration performed . Also the necessity for re calibration if needed. The factors that influence/govern this are sensor mounting, sensor data quality and cross calibration features. Certain metrics can be defined which govern these. However nothing exists today that is standardized. 
Camera calibration\cite{b9} is a very important step in computer vision applications. It means the computation of intrinsic and extrinsic parameters of a camera, which are used for the correction of distortions and the mapping of the 3D world points to the 2D image points. This tutorial is intended to cover the main concepts and techniques pertaining to camera calibration.

\begin{itemize}
    \item Intrinsic parameters: These describe information regarding the internal camera structure, such as the focal length, principal point, and radial and tangential distortion coefficients.
\item Extrinsic parameters: These refer to the camera's external orientation, detailing its rotations and translations.
\item Calibration patterns: Images used in the capture of calibration images. Examples are checkerboards.
\item Calibration software: Specialized software tools can automate the calibration process with an accurate estimation of parameters.
\end{itemize}
Proper application of techniques accompanied by knowledge will make your computer vision applications more accurate and reliable.
\section{Vehicle in Loop}
Vehicle-in-the-loop (VIL) testing is an essential part of the development and validation process of autonomous driving systems. It involves the incorporation of a simulated environment and a real-world vehicle, in conjunction with control inputs, in order to test the vehicle against several driving scenarios.

\subsection{Major constituents of a VIL test setup}
\begin{itemize}
\item Real vehicle : Physical vehicle which is completely equipped with sensors, actuators, and autonomous driving system.
\item Simulator: A software-based simulation environment\cite{b10} that mimics real-life driving by including roads and conditions, traffic, and other vehicles in the driving scenarios.
\item Interface: Hardware and software interface that links the real vehicle with the simulator where real-time exchange and control data happens.
\end{itemize}
\subsection{VIL testing process}
\begin{itemize}
\item Scenario Generation: The simulator creates different driving scenarios ranging from straightforward highway driving to high complexity in intersections with heavy traffic patterns.
\item Sensor Simulation: The simulator provides realistic sensor data, which is derived from a simulated environment, including the camera image, LiDAR point cloud, and radar data.
\item Vehicle Response: The Vehicle's sensors perceive the simulated environment through the interface, and the autonomous driving system processes the data to create decisions on how to navigate.
\item Vehicle Action: The vehicle actuators, such as steering, acceleration and braking are therefore executed according to the autonomous driving system commands.
\item Simulation Update: The simulator updates the virtual environment based on the behavior of the vehicle and sensor measurements .
\item Loop: The process repeats again, forming a form of persistent interaction between the physical vehicle and the simulated environment.
\end{itemize}
\subsection{Benefits of VIL testing}
\begin{itemize}
\item Validation in Real World: VIL testing offers an opportunity of estimating how well the system can perform in real world with natural factors such as sensor noise, environmental interference, and unexpected events.
\item Accelerated testing: VIL testing simulates various conditions that can be far more rapid for development and testing purposes compared to real-world testing.\cite{b11}
\item Safety: VIL testing can be done under controlled conditions with no danger of accidents and, therefore, injury.
\item Cost-effective: Cost-wise, VIL testing can be relatively less costly than real-world testing in that it does away with the need for extensive road testing and utilization of roads.
\end{itemize}

VIL testing is very important because it combines the strengths of simulation and real-world testing, which are crucial in the pursuit of ensuring safety and reliability for autonomous vehicles.
\section{Software in Loop}
SIL testing is one of the key techniques in developing autonomous driving systems. It tests software parts in a simulated environment, but does not require actual hardware. Because it enables rapid and inexpensive evaluation of software functionalities and algorithms, it is possible.

\subsection{How SIL Testing Works}
\begin{itemize}
\item Software Development: Develop and compile software components, including perception algorithms and motion planning modules together with control algorithms.
\item Simulation Environment: Create a virtual environment that simulates real-world driving scenarios, including road conditions, traffic, and other vehicles.
\item Software Integration: Integrate the software components into the simulation environment.
\item Test Execution: Run various test cases to evaluate the software's performance under different conditions.
\item Data Analysis: Analyze the simulation results to identify potential issues and areas for improvement\cite{b12}.
\end{itemize}
\subsection{Benefits of SIL Testing}
\begin{itemize}
\item Early Bug Detection: Identify and fix software bugs early in the development process.
\item Accelerated Development: Develop a development cycle much faster as it reduces the overall time for physical testing.
\item Cost-Effective: Saves money by not having to use costly hardware and pricey test environments.
\item Safe Testing: Test it extensively without a risk of physical damage or injury.
\item Flexibility: The simulation environment can be easily changed and re-configured to reflect various scenarios.
\end{itemize}

\subsection{Key Applications of SIL Testing for Autonomous Driving}
\begin{itemize}
\item Algorithm verification-the accuracy and performance test of perception algorithms, motion planning algorithms, and control algorithms.
\item Sensor fusion-merging data from a variety of sensors (cameras, LiDAR, radar, etc.) into a single system for better perception
\item Functional safety test-safety-critical function analyses of the autonomous car
\item Performance optimization-parameter optimization in algorithms for better performance.
\end{itemize}
This enables engineers to improve the quality and reliability of autonomous driving systems using an ample amount of SIL testing.
\section{Hardware in Loop}

Hardware-in-the-Loop (HIL) testing is an essential technique in the development of autonomous driving systems. It ensures that the real-world component, such as an ECU, can be tested in a simulated environment. This bridges the gap between software simulation and real-time testing.

\subsection{Main Components of a HIL Test Setup}
\begin{itemize}
\item Real World Component: This is the actual, to-be-tested, physical component (ECU).
\item Simulator: A simulation software environment that produces realistic sensor data eg, camera images, LiDAR point clouds, radar data and vehicle dynamics
Real-Time Interface: Hardware interface connecting the real component to the simulator. This enables real-time communication and data exchange.
\item Simulation : The simulator generates a realistic driving scenario including road conditions, traffic and other vehicles.
\item Sensor Simulation: The simulator generates simulated sensor data using the virtual environment. It is fed into the ECU.
\item Response of ECU: Simulated sensor data is processed by the ECU and control signals are developed.
\item Actuator Simulation: The actuator simulator develops, depending on the ECU-issued control signals, the responses of the actuators, like steering, braking, acceleration.
\item Feedback Loop: The actuator simulator's responses are fed back to the ECU and a closed-loop system is formed\cite{b13}.
\end{itemize}
\subsection{Benefits of HIL Testing}

\begin{itemize}
\item Early Validation: Validate elements at the beginning of the development cycle.
Fast Development: Fast development cycle with testing of multiple cases in one go.
\item Economical: Economical compared to actual testing, especially if cases are rare and fatal.
\item Safe Testing: Testing of elements in a controlled environment without material damage or injury.
\item High Test Coverage: Simulation of extreme driving scenarios the vehicle is going to go through.
\end{itemize}
HIL combines the benefits of simulation with real-world testing into one model and plays a very significant role in ascertaining the safety and reliability of autonomous driving systems.

\section{Final Product Acceptance Test(FPAT)}
The needed safety, reliability, and performance characteristics must be met and assured before the deployment of the autonomous vehicle. This phase embraces a holistic approach in the evaluation of the AV's hardware, software, and the system in its entirety.

\subsection {FPAT Components}

\subsubsection{Road Tests}

Testing the AV in a range of realistic conditions (urban, highway, and rural roads) to test the AV's capability in changing scenarios.
Sensor performance, perception algorithms, and decision-making capabilities will be assessed.
Response of AV during unexpected and edge-case situations
Scenario-based testing will be carried out.
Performance simulation testing with the AV on specific scenarios such as lane changes, intersections, and emergency braking.
Testing complex traffic scenarios and adverse weather conditions using the AV.

\subsubsection{Functional safety tests}
Checks on the adherences of the AV to functional safety standards for standards (e.g., ISO 26262)\cite{b14}.
Checking the mechanism of fault tolerance together with redundancy measures.
Capability assessment of AV in terms of its ability to manage failures inside the systems for safe operations.
\subsubsection{Cybersecurity Tests}

Analysis to identify the vulnerabilities of AVs towards cyberattacks and data breaches
Testing the security of the software and communication of the AV systems from attacks
Regulatory Compliance Tests:

Verification regarding the compliance with all the relevant regulations and standards, (SAE Level 4/5 standards)
Its verification with the requirements for license and its deployment.
FPAT will confirm that the AV must meet and ensure any applicable safety, performance, and regulatory requirements before the AV is deployed on public roads.
\section{Cost issues for testing AVs}
The expense, economies of scale, and inefficient features of testing prototypes of autonomous vehicles on the real physical environment remain a problem.

\subsection{High Costs and Resource Constraints}

Producing the prototypes is labor and cost-intensive.
Development and maintenance of roads, traffic environments, and their safety features is expensive.Complying with regulations and obtaining the necessary permits for testing roads is complicated and time-consuming.
There is a limitation on the number of prototypes that can be tested at a given time.It is tough and resource intensive to try and generate a very wide and broad range of diverse and challenging driving scenarios in the real world.Accidents and collisions during testing often result in massive damage and delays.A lot of time and resources are wasted with every iteration of the physical prototype. Thus, the pace of development is slowed down.

\subsection{Overcome Challenges}

The autonomous vehicle industry has become highly reliant on simulation-based testing to overcome the challenges. Multiple driving scenarios could be simulated at once, which would quicken the cycle of development, reduce the cost, and ensure safety.

Simulation benefits by testing multiple scenarios.Total control is present over the test environment.There is no possibility of physical damage.It saves costs related to physical prototypes and testing infrastructure. Developers will come up with safer, more reliable efficient autonomous vehicles by using physical-virtual testing.

One of the large-scale development issues with Autonomous Vehicles is the ODD problem; the problem refers to a wide variety of different driving scenarios that an AV must cope with, and therefore cannot be fully simulated within a real-world test.

There are immense problems with traditional approaches.
The cost of physical testing is enormous while also requiring a tremendous amount of resources for vehicle manufacturing, testing infrastructure, and personnel.Simulation only allows testing of a few vehicles and test scenarios.Physics-based testing always involves some risk of accidents and damage.
Physical testing-based iterative development cycles can be very time-consuming and inefficient.

\subsection{Model-based design as a solution}

Model-based design promisingly solves the ODD problem because engineers can develop virtual models of vehicles and their environments and simulate wide ranges of scenarios by testing different designs iterations in efficient manners.\cite{b15}

Key benefits of model-based design:

\begin{itemize}
\item Cost-effective: This is because there will be lesser reliance on physical prototypes and testing infrastructures.
\item Accelerated development: Also, the cycle of iteration will be faster, and the appearance of design flaws will be quicker.
\item Increased safety: Also, critical systems can then be tested in a virtual setting.
\item Scalability: It can simulate millions of scenarios while still testing out edge cases.
\end{itemize}
With the advent of model-based design, the automotive industry can really address the problem of ODD through the accelerating design of safe and reliable autonomous vehicles.

\section{Future Scope}

The future of verification and validation for autonomous systems will likely focus on addressing the growing complexity of AI integration, ensuring reliability, and maintaining safety in dynamic and unpredictable environments.\cite{b16} Here are some key trends shaping verification and validation processes for these systems.

\subsection{AI based Verification and Validation}
  AI-powered simulators will reveal complex real-world scenarios, including rare edge cases, for stress-testing autonomous systems. Adaptive AI algorithms will dynamically refine test scripts based on observed system performance. Additionally, automated fault detection using AI models will enhance error detection and failure prediction by identifying subtle inconsistencies in behavior and anticipating potential failure modes with greater accuracy. These AI-driven advancements will play a pivotal role in strengthening the verification and validation processes for autonomous systems, ensuring their safety and reliability.

\subsection{Scenario-Based Testing and Coverage :}
 Comprehensive scenario libraries, encompassing both normal and edge cases, will be critical for ensuring thorough testing of autonomous systems. Industry-standard testing scenarios will ensure consistency across manufacturers and platforms, supported by standardized benchmarks and key performance indicators (KPIs) to evaluate software and system quality and safety. These measures will enable more consistent and robust testing across autonomous driving systems. Additionally, the development of formal methods for verifying safety-critical software and system components will be essential. This includes techniques for formally specifying system and software requirements and validating them through formal methods to enhance the robustness and reliability of autonomous vehicle systems.

\subsection{Combining Physical and Virtual Testing:}
 A hybrid approach that integrates virtual simulations with real-world testing provides robust validation for complex automotive systems. Virtual testing extends the range of scenarios tested, including edge cases and hazardous conditions that are difficult or unsafe to replicate physically. Meanwhile, real-world testing ensures the system performs reliably in actual operational environments, bridging the gap between theoretical validation and practical application. Together, these methods complement each other, addressing their respective limitations and strengthening the overall testing process.

\subsection{Real-World Data Feedback :}
Real-World Data Feedback: Continuous integration of real-world operational data into the verification and validation process enhances the precision and accuracy of testing. By incorporating real-world insights, this approach helps uncover unforeseen issues, ensures adaptability by introducing new scenarios, and aligns system behavior more closely with actual operating conditions. This iterative feedback loop refines the testing process, improving system reliability and safety over time.

\section{Conclusion}
Developing safe and reliable software for autonomous vehicles presents significant challenges due to the complexity of these systems and the diverse range of operational scenarios they must handle. This paper outlines key strategies for preventing, detecting, and mitigating software defects throughout the autonomous vehicle development lifecycle.

The V-model software development process provides a structured framework for verification and validation activities. Simulation plays a critical role in testing autonomous vehicle software, enabling accelerated development and evaluation across a wide range of scenarios. Hardware-in-the-loop (HIL) and vehicle-in-the-loop (VIL) testing bridge the gap between simulation and real-world testing, ensuring robust performance under realistic conditions.

Rigorous calibration, extensive road testing, and comprehensive final product acceptance testing are essential to ensure autonomous vehicles meet stringent safety and performance requirements before deployment. However, the high costs and resource constraints of physical testing remain considerable challenges.

Model-based design and the increased use of simulation offer promising solutions for addressing the challenges of the operational design domain and enabling the efficient development of safe autonomous vehicles. As technology advances, ongoing research and development of improved testing methodologies, tools, and standards will be crucial to realizing the potential of autonomous vehicles while maintaining public safety.

By leveraging the approaches outlined in this paper and continuing to advance the state-of-the-art in autonomous vehicle software development, the industry can progress toward producing highly reliable self-driving systems capable of operating safely in complex and dynamic environments. Nonetheless, significant work remains before fully autonomous vehicles can be widely deployed with confidence.

\end{document}